\documentstyle[12pt]{article}
\newcommand{\bib}{\bibitem}
\newcommand{\be}{\begin{equation}}
\newcommand{\ee}{\end{equation}}
\newcommand{\er}{\end{eqnarray}}
\newcommand{\br}{\begin{eqnarray}}

\begin{document}
\thispagestyle{empty}
\begin{raggedleft}
hep-th/9710060\\
September/97\\
\end{raggedleft}
$\phantom{x}$\vskip 0.618cm\par
{\huge \begin{center} Duality Symmetry and Soldering in Different Dimensions
\end{center}}\par

\begin{center}
$\phantom{X}$\\
{\Large R.Banerjee\footnote{On leave
of absence from S.N.Bose National Centre for Basic
Sciences, Calcutta, India. e-mail:rabin@if.ufrj.br} and C.Wotzasek}\\[3ex]
{\em Instituto de F\'\i sica\\
Universidade Federal do Rio de Janeiro\\
21945, Rio de Janeiro, Brazil\\}
\end{center}\par
\begin{abstract}
We develop a systematic method of obtaining duality symmetric actions in
different dimensions.
This technique is applied for the quantum mechanical harmonic
oscillator, the scalar 
field theory in two dimensions and the Maxwell theory in four dimensions. 
In all cases there are two such distinct actions. Furthermore, by 
soldering these distinct
actions in any dimension a master action 
is obtained which is duality invariant under a
much bigger set of symmetries than is usually envisaged. The concept of
swapping duality is introduced and its implications are discussed. 
The effects of coupling to gravity are also elaborated. Finally, the
extension of the analysis for arbitrary dimensions is indicated.
\end{abstract}


\newpage

\section{Introduction}
The intriguing role of duality in different contexts is being
progressively understood and clarified \cite{O, GH, JS, AG}. 
Much effort has been given
in sorting out several technical aspects of duality symmetric actions.
In this context the old idea \cite{O, Z, DT} of eletromagnetic duality has been
revived with considerable attention and emphasis \cite{SS, GR, NB, DGHT}. 
Recent directions \cite{SS, KP, PST, G} also
include an 
abstraction of manifestly covariant forms  for such
actions or an explicit proof of their equivalence with the nonduality
symmetric actions, which they are supposed to represent. There are also
different suggestions on the possible analogies between duality symmetric
actions in different dimensions. In particular it has been claimed
\cite{PST} that
the two dimensional self dual 
action given in \cite{FJ} is the analogue of the four
dimensional electromagnetic duality symmetric action \cite{SS}.
In spite of the recent spate of papers on this subject there does not
seem to be a simple clear cut way of arriving at duality symmetric actions.
Consequently the fundamental nature of duality remains clouded by
technicalities. Additionally, the dimensionality of space time appears to
be extremely crucial. For instance, while the 
duality symmetry in $D=4k$ dimensions
is characterised by the one-parameter continuous group $SO(2)$, that in
$D=4k+2$ dimensions is described by a discrete group with just two
elements \cite{DGHT}.
Likewise, it has also been argued from general notions that a symmetry 
generator exists only in the former case. From an algebraic point of view
the distinction between the dimensionalities is manifested by the
following identities,
\br
\label{i1}
\mbox{}^{**}F &=& F\,\,\,;\,\,\,D=4k+2\nonumber\\
 &=&- F\,\,\,;\,\,\,D=4k
\er
where the $*$ denotes a usual Hodge 
dual operation and $F$ is the $\frac{D}{2}$-form.
Thus there is a self dual operation in the former which is missing in the
latter dimensions. This apparently leads to separate consequences for duality
in these cases.

The object of this paper is to develop a method for systematically
obtaining and investigating different aspects of
duality symmetric actions that embrace all dimensions. A deep unifying
structure is illuminated which also leads to new symmetries. Indeed we
show that duality is not limited to field or string theories, but is
present even in the simplest of quantum mechanical examples- the harmonic
oscillator. It is precisely this duality which pervades all field
theoretical examples as will be explicitly shown. The basic idea of our
approach is deceptively simple. We start from the second order action for
any theory and
convert it to the first order form by introducing an auxiliary variable.
Next, a suitable relabelling of variables is done which induces an
internal index in the theory.  It is crucial to note that 
there are two distinct classes of
relabelling characterised by the opposite signatures of the
determinant of the $2\times 2$ orthogonal matrix defined in the internal 
space. 
Correspondingly, in this  space there are two  actions that are 
manifestly duality symmetric. 
Interestingly, their equations of motion are just the self and anti-self
dual solutions, where the dual field in the internal space 
is defined below in (\ref{i2}). It
is also found that in all cases there is one (conventional duality) 
symmetry transformation which preserves the invariance of these actions
but there is another transformation which swaps the actions. We refer
to this property as swapping duality. This
indicates the possibility, in any dimensions, 
of combining the two actions to a master action
that would contain all the duality symmetries.
Indeed this construction is explicitly done by
exploiting the ideas of soldering introduced in \cite{S} and developed by
us \cite{ABW, BW}. The soldered master action also  has manifest Lorentz
or general coordinate invariance. The generators of the symmetry
transformations are also obtained. 

It is easy to visualise how the internal space effectively 
unifies the results in the different $4k+2$
and $4k$ dimensions. The dual field is now defined to include the internal
index $(\alpha, \beta)$ in the fashion,
\br
\label{i2}
\tilde F^\alpha &=&\epsilon^{\alpha\beta}\mbox{}^{*}F^\beta 
\,\,\,;\,\,\,D=4k\nonumber\\
\tilde F^\alpha &=&\sigma_1^{\alpha\beta}\mbox{}^{*}F^\beta 
\,\,\,;\,\,\,D=4k+2
\er
where $\sigma_1$ is the usual Pauli matrix and 
$\epsilon_{\alpha\beta}$ is the fully
antisymmetric $2\times 2$ matrix with $\epsilon_{12} =1$. Now, 
irrespective of the
dimensionality, the repetition of the dual operation yields,
\be
\label{i3}
\tilde{\tilde F} = F
\ee
which generalises the relation (\ref{i1}). An immediate consequence of
this is the possibility to obtain self and anti-self dual solutions in
all even $D=2k+2$ dimensions. Their explicit realisation is one of
the central results of the paper.

The paper is organised into five sections. In section 2 the above ideas
are exposed by considering the example of the simple
harmonic oscillator. A close parallel with the electromagnetic notation is
also developed to illuminate the connection between this exercise and
those given for the field theoretical models in the next two sections. The
duality of scalar field theory in two dimensions is considered in section 3.
The occurrence of a pair of actions is shown which exhibit duality and
swapping symmetries. These are the analogues of the four
dimensional electromagnetic duality symmetric actions.
Indeed, from these expressions,it is a trivial matter to
reproduce both the self and anti-self 
dual actions given in \cite{FJ}. Our analysis clarifies several issues
regarding the intertwining roles of chirality and duality in two dimensions.
The soldering of the pair of duality symmetric actions 
is also performed leading to fresh insights.
The analysis is completed by
including the effects of gravity. In section 4, the Maxwell theory is
treated in great details. Following our prescription the duality symmetric
action given in \cite{SS} is obtained. However, there is also a new
action which is  duality symmetric. Once again the soldering of these actions
leads to a master action which contains a much richer structure of
symmetries. Incidentally, it also manifests the original symmetry
that interchanges the Maxwell equations with the Bianchi identity, but
reverses the signature of the action.  As usual, the effects of gravity are
straightforwardly included. Section 5 contains the concluding comments.

\section {Duality in $0+1$ dimension}
The basic features of duality symmetric actions are already present
in the quantum mechanical examples as the present analysis on the
harmonic oscillator will clearly demonstrate. Indeed, this simple 
example is worked out in some details to illustrate the key concepts of
our approach and set the general tone of the paper. An extension to field
theory is more a matter of technique rather than introducing truly new
concepts. The
Lagrangean for the one-dimensional oscillator is given by,
\be
L=\frac{1}{2}\Big ({\dot q}^2-q^2\Big)
\label{10}
\ee
leading to an equation of motion,
\be
\ddot q+q=0
\label{20}
\ee
Introducing a change of variables,
\be
E=\dot q\,\,\,\,\,; \,\,\,\,\, B=q
\label{30}
\ee
so that,
\be
\dot B-E=0
\label{40}
\ee
is identically satisfied, the above equations (\ref{10}) and (\ref{20}) 
are, respectively, expressed as follows;
\be
L=\frac{1}{2}\Big(E^2-B^2\Big)
\label{50}
\ee
and,
\be
\dot E+B=0
\label{60}
\ee
It is simple to observe that  the transformations,\footnote{Note that
these are just the discrete cases ($\alpha=\pm\frac{\pi}{2}$) for a general
$SO(2)$ rotation matrix parametrised by the angle $\alpha$}
\be
E\rightarrow \pm B\,\,\,;\,\,\, B\rightarrow \mp E
\label{70}
\ee
swaps the equation of motion (\ref{60}) with the identity (\ref{40})
although the Lagrangean (\ref{50}) is not invariant. The similarity
with the corresponding analysis in the Maxwell theory is quite striking,
with $q$ and $\dot q$ simulating the roles of the magnetic and electric
fields, respectively. There is a duality among the equation of motion
and the `Bianchi' identity (\ref{40}), which is not
manifested in the Lagrangean.

In order to elevate the duality to the Lagrangean, the basic step is to
rewrite (\ref{10}) in the first order form by introducing an
additional variable,
\br
L&=&p\dot q-\frac{1}{2}(p^2+q^2)\nonumber\\
&=&\frac{1}{2}\Big(p\dot q-q\dot p -p^2-q^2\Big)
\label{80}
\er
where a symmetrisation has been performed. There are now
two possible classes for relabelling these variables corresponding to
proper and improper rotations generated by the matrices $R^+(\theta)$ and
$R^-(\varphi)$ with determinant $+1$ and $-1$, respectively,
\br
\left(\begin{array}{c}
q\\
p
\end{array}\right)
=
\left(\begin{array}{cc}
{\cos\theta} & {\sin\theta} \\
{-\sin\theta} &{\cos\theta} \end{array}\right)
\left(\begin{array}{c}
x_1\\
x_2
\end{array}\right)
\label{matrix1}
\er
\br
\left(\begin{array}{c}
q\\
p
\end{array}\right)
=
\left(\begin{array}{cc}
{\sin\varphi} & {\cos\varphi} \\
{\cos\varphi} &{-\sin\varphi} \end{array}\right)
\left(\begin{array}{c}
x_1\\
x_2
\end{array}\right)
\label{matrix}
\er
leading to the distinct Lagrangeans,
\br
L_\pm&=&\frac{1}{2}\Big(\pm x_\alpha\epsilon_{\alpha\beta}\dot x_\beta
-x_\alpha^2\Big)\nonumber\\
&=&{1\over 2}\left( \pm B_\alpha\epsilon_{\alpha\beta}E_\beta-B_\alpha^2
\right)
\label{100}
\er
where we have reverted back to the `electromagnetic' notation
introduced in (\ref{30}).  By these change
of variables an index $\alpha=(1, 2)$ has been introduced that
characterises a symmetry in this internal space, the complete details of
which will progressively become clear.
It is useful to remark that the above change of variables are succinctly
expressed as,
\br
q&=&x_1\,\,\,\,;\,\,\,\, p=x_2\nonumber\\
q&=&x_2\,\,\,\,;\,\,\,\, p=x_1
\label{90}
\er
by setting the angle $\theta=0$ or $\varphi =0$ in the rotation matrices 
(\ref{matrix1}) and (\ref{matrix}). Correspondingly,
the Lagrangean (\ref{80}) goes over to (\ref{100}). 
Now observe
that the above Lagrangeans (\ref{100}) 
are manifestly invariant under the continuous duality
transformations,
\be
\label{90a}
x_\alpha\rightarrow R^+_{\alpha\beta}x_\beta
\ee
which may be equivalently expressed as,
\br
E_\alpha&\rightarrow& R^+_{\alpha\beta}E_\beta\nonumber\\
B_\alpha&\rightarrow& R^+_{\alpha\beta}B_\beta
\label{100a}
\er
where $R^+_{\alpha\beta}$ is the usual $SO(2)$ rotation matrix (\ref{matrix1}).
The generator of the infinitesimal symmetry transformation is given by,
\be
\label{gen}
Q^\pm =\pm\frac{1}{2}x_\alpha x_\alpha
\ee
so that the complete transformations (\ref{90a}) are generated as,
\br
x_\alpha \rightarrow x'_\alpha & =& e^{-i\theta Q} x_\alpha e^{i\theta Q}
\nonumber\\ 
&=& R_{\alpha\beta}^+(\theta)x_\beta
\label{fingen}
\er
This is easy to verify by using the basic symplectic brackets obtained
from (\ref{100}),
\be
\label{symplectic}
\{x_\alpha , x_\beta\}=\mp \epsilon_{\alpha\beta}
\ee
Parametrising the angle by $\theta=\frac{\pi}{2}$ the discrete
transformation is obtained,
\br
E_\alpha&\rightarrow& \epsilon_{\alpha\beta}E_\beta\nonumber\\
B_\alpha&\rightarrow& \epsilon_{\alpha\beta}B_\beta
\label{110}
\er
This is the parallel of the usual constructions 
done in the Maxwell theory to
induce a duality symmetry in the action. 

Let us now comment on an
interesting property, which is related to the existence of two distinct
Lagrangeans (\ref{100}), by replacing (\ref{100a}) with a new set of
transformations,
\br
E_\alpha&\rightarrow& R^-_{\alpha\beta}(\varphi)E_\beta\nonumber\\
B_\alpha&\rightarrow& R^-_{\alpha\beta}(\varphi)B_\beta
\label{110b}
\er
Notice that these transformations preserve the invariance of
the Hamiltonian following from either $L_+$ or $L_-$. 
Interestingly, the kinetic
terms in the Lagrangeans change signatures so that $L_+$ and $L_-$ 
swap into one another. This feature of duality swapping
will subsequently recur in a different context and has important
implications in higher dimensions.

The discretised version of (\ref{110b}) is obtained by setting 
$\varphi =0$,
\br
E_\alpha&\rightarrow& \sigma_1^{\alpha\beta}E_\beta\nonumber\\
B_\alpha&\rightarrow& \sigma_1^{\alpha\beta}B_\beta
\label{110a}
\er
It is precisely the $\sigma_1$ matrix that reflects the proper into
improper rotations,
\be
\label{everton}
R^+(\theta) \sigma_1=R^-(\theta)
\ee
which illuminates the reason behind the swapping of the Lagrangeans in
this example.

Since we have systematically developed a procedure for obtaining a duality
symmetric Lagrangean, it is really not necessary to show its equivalence
with the original Lagrangean, as was done in the Maxwell theory. 
Nevertheless, to complete the analogy, we show that (\ref{100}) 
reduces to (\ref{50}) or (\ref{10}) by using the equation of motion,
\be
x_\alpha=\pm\epsilon_{\alpha\beta}\dot x_\beta
\label{120}
\ee
which can be reexpressed as,
\be
B_\alpha=\pm\epsilon_{\alpha\beta} E_\beta
\label{130}
\ee
to eliminate one component (say the variables with label 2) from (\ref{100}).
This immediately reproduces (\ref{50}) while (\ref{110}) reduces to (\ref{70}).

An important point to stress is that there are actually two, and not one,
duality symmetric actions $L_\pm$ (\ref{100}), corresponding to the
signatures in the determinant of the transformation matrices.  
As shall be shown in 
subsequent sections this is also true for the scalar field theory in $1+1$
dimensions and the electromagnetic theory in $3+1$ dimensions. Usually, in
the literature, only one of these 
is highlighted while the other is not mentioned.
We now elaborate the implications of this property which will also be
crucial in discussing field theoretical models. In the coordinate
language these Lagrangeans correspond to two bi-dimensional chiral
oscillators rotating
in opposite directions. This is easily
verified either by looking at the classical equations of motion or by
examining the spectrum of the angular momentum operator,
\be
J_\pm=\pm\epsilon_{ij}x_i p_j = \pm H
\label{140}
\ee
where $H$ is the Hamiltonian of the usual harmonic oscillator.
In other words the two Lagrangeans manifest the dual aspects of rotational
symmetry in the two-dimensional internal space. 
Consequently it is possible to solder them by following the
general techniques elaborated in \cite{ABW, BW}. This soldering as well as
its implications are the subject of the remainder of this section.

The soldering mechansim, it must be recalled, is intrinsically an operation
that has no classical analogue. The crucial point is that the Lagrangeans
(\ref{100}) are now considered as functions of independent variables,
namely $L_+(x)$ and $L_-(y)$, instead of the same $x$. A naive
addition of the classical Lagrangeans with the same variable is of course
possible leading to a trivial result. If,
on the other hand, the Lagrangeans are functions of distinct variables,
a straightforward addition does not lead to any new information. The soldering
process precisely achieves this purpose. Consider the gauging of the
Lagrangeans under the following gauge transformations,
\be
\delta x_\alpha=\delta y_\alpha=\dot\eta_\alpha
\label{150}
\ee
Then the gauge variations are given by,
\be
\delta L_\pm(z)=\epsilon_{\alpha\beta}
\dot\eta_\alpha J_\beta^{(\pm)}(z)\,\,\,\,;\,\,
z=x,y
\label{160}
\ee
where the currents are defined by,
\be
\label{170}
J_\alpha^{(\pm)}(z)=\pm\dot z_\alpha+\epsilon_{\alpha\beta}z_\beta
\ee
Introducing a new field $B_\alpha$ transforming as,
\be
\delta B_\alpha =\epsilon_{\beta\alpha}\dot\eta_\beta
\label{180}
\ee
that will effect the soldering, it
is possible to construct a first iterated Lagrangean,
\be
\label{190}
L_\pm^{(1)}=L_\pm-B_\alpha J_\alpha^\pm
\ee
The gauge variation of (\ref{190}) is easily obtained,
\be
\label{200}
\delta L_\pm^{(1)}=-B_\alpha\delta J_\alpha^\pm
\ee
Using the above results we define a second iterated Lagrangean,
\be
\label{210}
L_\pm^{(2)}=L_\pm^{(1)}-\frac{1}{2}B^2_\alpha
\ee
which finally leads to a Lagrangean,
\be
\label{220}
L=L_+^{(2)}(x)+L_-^{(2)}(y)=L_+(x) + L_-(y) -B_\alpha\Big(J_\alpha^+ (x)+
J_\alpha^- (y)\Big)-B_\alpha^2
\ee
that is invariant under the complete set of transformations
(\ref{150}) and (\ref{180}),i.e.;
\be
\label{230}
\delta L=0
\ee
It is now possible to eliminate the auxiliary $B_\alpha$ field by using
the equation of motion, which yields,
\be
\label{240}
B_\alpha=-\frac{1}{2}\Big (J_\alpha^+(x)+J_\alpha^-(y)\Big)
\ee
Inserting this solution back into (\ref{220}), we obtain the final
soldered Lagrangean,
\be
\label{250}
L(w)=\frac{1}{4}\Big(\dot w_\alpha^2-w_\alpha^2\Big)
\ee
which is no longer a function of $x$ and $y$ independently, but only
on their
gauge invariant combination,
\be
\label{260}
w_\alpha=\left(x_\alpha-y_\alpha\right)
\ee
The soldered Lagrangean just corresponds to a simple bi-dimensional 
oscillator. Thus, by starting from two Lagrangeans which contained the
opposite aspects of a duality symmetry, it is feasible to combine them
into a single Lagrangean which has a richer symmetry. A similar
phenomenon also exists in the field theoretical examples, as shall be
shown subsequently. 

Let us now expose all the symmetries of the above
Lagrangean. It is most economically done by recasting this Lagrangean in
two equivalent forms,
\be
L=\Omega_\alpha^+\Omega_\alpha^- = \bar\Omega_\alpha^+\bar\Omega_\alpha^-
\label{omega}
\ee
where,
\br
\Omega_\alpha^\pm &=&\frac{1}{2} \Big(
\dot w_\alpha\pm\Lambda_{\alpha\beta}w_\beta\Big)\nonumber\\
\bar\Omega_\alpha^\pm &=&\frac{1}{2} \Big(\Lambda_{\alpha\beta}\dot w_\beta
\pm w_\alpha\Big)\nonumber\\
\Lambda_{\alpha\beta}&=&\Big(R_{\alpha\beta}^+\,\,,\,\,
R_{\alpha\beta}^- \Big)
\label{lambda}
\er
Now the Lagrangean (\ref{omega}) is {\it manifestly} symmetric under the
following continuous dual transformations,
\be
w_\alpha\rightarrow R^\pm_{\alpha\beta}\,w_\beta
\label{dual}
\ee
The transformation involving $R^+$ is just the original symmetry
(\ref{100a}). Those involving the $R^-$ matrices are the new symmetries.
Recall that the latter transformations 
swapped the two independent Lagrangeans
$L_\pm$. The soldered Lagrangean contains both combinations and hence
manifests both these symmetries. The corresponding symmetry group is
therefore $O(2)$. This is a completely new phenomenon. It
also occurs in field theory with certain additional subtle features.

The generator of the infinitesimal transformations that leads to the
$SO(2)$ rotation in
(\ref{dual}) is given by,
\be 
\label{dualgen}
Q=w_\alpha \epsilon_{\alpha\beta} \pi_\beta
\ee
so that,
\br
\label{dualfingen}
w_\alpha\rightarrow w'_\alpha &=& e^{-i\theta Q}\, w_\alpha\, e^{i\theta
Q}\nonumber\\ 
&=& R^+_{\alpha\beta}(\theta)\, w_\beta
\er
which is verified by using the canonical brackets,
\be
\label{brackets}
\{w_\alpha, \pi_\beta\}=\delta_{\alpha\beta}
\ee

It is worthwhile to point out the quantum nature of the above calculation
by rewriting (\ref{250}), after an appropriate scaling of variables,  
in the form of an identity,
\br
\label{270}
L(x-y)&=&L(x)+L(y)-2x_\alpha^+y_\alpha^-\nonumber\\
z_\alpha^\pm&=&\frac{1}{\sqrt 2}(\dot z_\alpha\pm \epsilon_{\alpha\beta}
z_\beta)
\er
This shows that the Lagrangean of the 
simple harmonic oscillator expressed in terms of
the ``gauge invariant" variables $w=x-y$ is not obtained by just adding the
independent contributions. Rather, there is a contact term which manifests
the quantum effect. Indeed, the above identity can be interpreted as the
analogue of the well known Polyakov-Weigman \cite{PW} identity in two
dimensional field theory. As our analysis shows, such identities will
always occur whenever dual aspects of some symmetry are being soldered
or fused to yield a composite picture, irrespective of the
dimensionality of space-time \cite{ABW}. 
In the Polyakov-Weigman case it was the
chiral symmetry whereas here it was the rotational symmetry. 

\section{The Scalar Theory in 1+1 Dimensions}

The ideas developed in the previous section are now implemented and
elaborated in $1+1$
dimensions. It is simple to realise that the scalar theory is a very
natural example. For instance, in these dimensions, there is no photon and
the Maxwell theory trivialises so that the electromagnetic field can be
identified with a scalar field. Thus all the results presented here can be
regarded as equally valid for the ``photon" field. 
Indeed the computations will also be presented in a very suggestive
notation which  illuminates the Maxwellian nature of the problem.
Consequently the
present analysis is an excellent footboard for diving into the actual
electromagnetic duality discussed in the next section. The effects of
gravity are easily included in our approach as shown in a separate subsection.

The Lagrangean for the free massless scalar field is given by,
\be
\label{w10}
{\cal L}=\frac{1}{2} \Big(\partial_\mu\phi\Big)^2
\ee
and the equation of motion reads,
\be
\label{w20}
\ddot\phi-\phi ''=0
\ee
where the dot and the prime denote derivatives with respect to time and
space components, respectively. Introduce, as before, a 
change of variables using electromagnetic symbols,
\be
E=\dot\phi\,\,\,\,;\,\,\,\, B=\phi '
\label{w30}
\ee
Obviously, $E$ and $B$ are not independent but constrained by the
identity, 
\be
\label{w40}
E'-\dot B=0
\ee
In these variables the equation of motion and the Lagrangean are expressed
as,
\br
\label{w50}
&\mbox{}&\dot E-B'=0\nonumber\\
&\mbox{}&{\cal L}=\frac{1}{2}\Big(E^2-B^2\Big)
\er
It is now easy to observe that the transformations,
\be
\label{w60}
E\rightarrow \pm B\,\,\,\,;\,\,\,\, B\rightarrow \pm E
\ee
display a duality between the equation of motion and the `Bianchi'-like
identity (\ref{w40}) but the Lagrangean changes its signature.
Note that there is a relative change in the signatures
of the duality transformations (\ref{70}) and (\ref{w60}), arising
basically from dimensional considerations. This symmetry coresponds to the
improper group of rotations.

To illuminate the close connection with the Maxwell formulation, we
introduce covariant and contravariant vectors with a Minkowskian metric
$g_{00}=-g_{11}=1$, 
\be
F_\mu=\partial_\mu\phi\,\,\,;\,\,\, F^\mu=\partial^\mu\phi
\label{w60a}
\ee
whose components are just the `electric' and
`magnetic' fields defined earlier,
\be
F_\mu=\Big(E, B\Big)\,\,\,\,;\,\,\,\, F^\mu=\Big(E,- B\Big)
\label{w60b}
\ee
Likewise, with the convention $\epsilon_{01}=1$, the dual field is defined,
\br
\mbox{}^* F_\mu&=&\epsilon_{\mu\nu}\partial^\nu\phi =\epsilon_{\mu\nu}F^\nu 
\nonumber\\
&=& \Big(-B, -E\Big)
\label{w60c}
\er
The equations of motion and the `Bianchi' identity are now expressed by
typical electrodynamical relations,
\br
\partial_\mu F^\mu&=&0\nonumber\\
\partial_\mu \mbox{}^* F^\mu&=&0
\label{w60d}
\er

To expose a
Lagrangean duality symmetry, the basic principle of our approach 
to convert the original second order form
(\ref{w50}) to its first order version and then invoke a relabelling of
variables to provide an internal index, is adopted. This is easily achieved
by first introducing an auxiliary field,
\be
\label{w70}
{\cal L}= PE-\frac{1}{2}P^2-\frac{1}{2}B^2
\ee
where $E$ and $B$ have already been defined. The following renaming of
variables corresponding to the proper and improper transformations (see
for instance (\ref{matrix1}) and (\ref{matrix}) or (\ref{90})) is used,
\br
\label{w80}
\phi&\rightarrow&\phi_1\nonumber\\
P&\rightarrow&\pm\phi_2'
\er
where we are just considering the discrete sets (\ref{90}) of the full
symmetry (\ref{matrix1}) and (\ref{matrix}). 
Then it is possible to recast (\ref{w70}) in the form,
\br
\label{w90} 
{\cal L}\rightarrow{\cal L}_\pm&=&\frac{1}{2}\Bigg[\pm
{\phi'}_\alpha\sigma_1^{\alpha\beta}\dot\phi_\beta -{\phi'}_\alpha^2\Bigg] 
\nonumber\\
&=&\frac{1}{2}\Bigg[\pm
B_\alpha\sigma_1^{\alpha\beta}E_\beta-B_\alpha^2\Bigg] 
\er
In the second line the
Lagrangean is expressed in terms of the electromagnetic variables.
This Lagrangean is
duality symmetric under the transformations of the basic scalar fields,
\be
\label{w100}
\phi_\alpha\rightarrow\sigma^{\alpha\beta}_1\phi_\beta
\ee
which, in the notation of $E$ and $B$, is given by,
\br
\label{w110}
B_\alpha &\rightarrow &\sigma^{\alpha\beta}_1B_\beta\nonumber\\
E_\alpha &\rightarrow &\sigma^{\alpha\beta}_1E_\beta
\er
It is quite interesting to observe that, contrary to the harmonic
oscillator example or the electromagnetic theory discussed in the 
next section, the transformation matrix in the $O(2)$ space is not
the epsilon, but rather a Pauli matrix. 
This result is in  agreement with that found from general algebraic
arguments \cite{SS, DGHT} which stated that for $d=4k+2$ dimensions
there is a discrete $\sigma_1$ symmetry. Observe that 
(\ref{w110}) is a manifestation of the
original duality (\ref{w60}) which was also effected by the same operation.
It is important to stress that the above symmetry is only
implementable at the discrete level. Moreover, since it is not connected
to the identity, there is no generator for this transformation.

To complete the picture, we also
mention that the following rotation,
\be
\phi_\alpha\rightarrow \epsilon_{\alpha\beta}\phi_\beta
\label{w120}
\ee
interchanges the Lagrangeans (\ref{w90}),
\be
{\cal L}_+\leftrightarrow {\cal L}_-
\label{w130}
\ee
Thus, except for a rearrangement of the the matrices generating the
various transformations, most features of
the simple harmonic oscillator example are perfectly retained. The crucial
point of departure is that now all these transformations are only discrete.
Interestingly, the master action constructed below lifts these symmetries
from the discrete to the continuous.

Let us therefore solder the two distinct Lagrangeans to
manifestly display the complete symmetries. Before doing this it is
instructive to unravel the self and anti-self dual aspects of these
Lagrangeans, which are essential to physically understand the soldering
process.  The equations of motion following from (\ref{w90}),  
in the language of the basic fields, is given by,
\be
\partial_\mu\phi_\alpha=\mp\sigma^1_{\alpha\beta}\epsilon_{\mu\nu}
\partial^\nu\phi_\beta
\label{dual1}
\ee
provided reasonable boundary conditions are assumed. Note that although
the duality symmetric Lagrangean is not manifestly Lorentz covariant, the
equations of motion possess this property. We will return to this aspect
again in the Maxwell theory.
In terms of a vector field $F_\mu^\alpha$ and its dual $\mbox{}^*
F_\mu^\alpha$ defined analogously to
(\ref{w60b}), (\ref{w60c}), the equation of motion is rewritten as,
\be
\label{motion}
F_\mu^\alpha=\pm\sigma_1^{\alpha\beta}\mbox{}^* F_\mu^\beta=\pm\tilde
F_\mu^\alpha
\ee
where the generalised Hodge dual $(\tilde F)$ has been defined in (\ref{i2}).
This explicitly reveals the self and anti-self dual nature of the
solutions in the combined internal and coordinate spaces. The result
can be extended to any $D=4k+2$ dimensions with suitable insertion of indices.

We now  solder the two Lagrangeans. 
This is best done by using the notation of the basic fields  
of the scalar theory. These Lagrangeans ${\cal L}_+$ and ${\cal L}_-$ 
are regarded as functions of the
independent scalar fields $\phi_\alpha$ and $\rho_\alpha$. 
Consider the gauging of the following symmetry,
\be
\delta\phi_\alpha =\delta\rho_\alpha=\eta_\alpha
\label{w140}
\ee
Following exactly the steps performed for the harmonic oscillator example
the final  Lagrangean analogous to (\ref{220}) is obtained,
\be
{\cal L}={\cal L}_+(\phi)+{\cal
L}_-(\rho)-B_\alpha\Big(J_\alpha^+(\phi) +J_\alpha^-(\rho)\Big)-B_\alpha^2
\label{w150}
\ee
where the currents are given by,
\be
J_\alpha^\pm(\theta)=\pm\sigma_{\alpha\beta}^1\dot\theta_\beta-{\theta'}_\alpha
\,\,\,;\,\,\,\theta=\phi\,\,,\,\,\rho
\label{w160}
\ee
The above Lagrangean is gauge invariant under the extended
transformations including (\ref{w140}) and,
\be
\delta B_\alpha =\eta_{\alpha}'
\label{w170}
\ee
Eliminating the auxiliary $B_\alpha$ field using the equations of motion,
the final soldered Lagrangean is obtained from (\ref{w150}),
\be
{\cal L}(\Phi)= \frac{1}{4}\partial_\mu\Phi_\alpha \partial^\mu\Phi_\alpha
\label{w180}
\ee
where, expectedly, this is now only a function of the gauge invariant
variable, 
\be
\Phi_\alpha=\phi_\alpha-\rho_\alpha
\label{w190}
\ee
This master Lagrangean possesses all the symmetries
that are expressed by the continuous
transformations, 
\be
\label{205}
\Phi_\alpha\rightarrow R^\pm_{\alpha\beta}(\theta)\Phi_\beta
\ee
The generator corresponding to the $SO(2)$ transformations is easily obtained,
\br
\label{206}
Q&=&\int dy \Phi_\alpha \epsilon_{\alpha\beta} \Pi_\beta\nonumber\\
\Phi_\alpha\rightarrow \Phi'_\alpha &=&e^{-i\theta Q} \Phi_\alpha e^{i\theta
Q}
\er
where $\Pi_\alpha$ is the momentum conjugate to $\Phi_\alpha$.
Observe that either the original symmetry in $\sigma_1$ or the swapping
transformations were only at the discrete
level. The process of soldering has lifted these
transformations from the discrete to the continuous form. It is equally
important to reemphasize that the master action now possesses the $SO(2)$
symmetry which is more commonly associated with four dimensional duality
symmetric actions, and not for two dimensional theories. Note that by
using the electromagnetic symbols, the Lagrangean can be 
displayed in a form which manifests the soldering effect of
the self and anti self dual symmetries (\ref{motion}),
\be
\label{207}
{\cal L}=\frac{1}{8}\Big(F_\mu^\alpha+\tilde F_\mu^\alpha\Big)
\Big(F^\mu_\alpha-\tilde F^\mu_\alpha\Big)
\ee
where the generalised Hodge dual in $D=4k+2$ dimensions has been
defined in (\ref{i2}).

An interesting observation is now made. Recall that the original duality
transformation (\ref{w60}) switching equations of motion into Bianchi
identities may be rephrased in the internal space by,
\br
\label{bianchi}
E_\alpha &\rightarrow& \mp R^\pm_{\alpha\beta} B_\beta\nonumber\\
B_\alpha &\rightarrow& \mp R^\pm_{\alpha\beta} E_\beta
\er
which is further written directly in terms of the scalar fields,
\be
\label{bianchi1}
\partial_\mu \Phi_\alpha \rightarrow \pm R^\pm_{\alpha\beta}
\epsilon_{\mu \nu}\partial^\nu \Phi_\beta
\ee
It is simple to verify that under these transformations even 
the Hamiltonian for the theories
described by the Lagreangeans
${\cal L}_\pm$ (\ref{w90}) are not invariant. However the Hamiltonian
following from the master
Lagrangean (\ref{w180}) preserves this symmetry. The Lagrangean itself
changes 
its signature. This is the exact analogue of the original situation.
A similar phenomenon also
occurs in the electromagnetic theory. This completes the discussion on the
symmetries of the master Lagrangean.

It is now straightforward to give a Polyakov-Weigman type identity,
that relates the ``gauge invariant" Lagrangean with the non gauge invariant
structures, by reformulating (\ref{w180}) after a scaling of the fields
$(\phi, \rho)\rightarrow \sqrt 2(\phi, \rho)$,
\be
\label{w210}
{\cal L}(\Phi)= {\cal L}(\phi)+{\cal L}(\rho)-2\partial_+\phi_\alpha
\partial_-\rho_\alpha
\ee
where the light cone variables are given by,
\be
\label{lc}
\partial_\pm =\frac{1}{\sqrt 2}(\partial_0 \pm \partial_1)
\ee

Observe that, as in the harmonic oscillator example, the gauge invariance
is with regard to the transformations introduced for the soldering of the
symmetries. Thus, even if the theory does not have a gauge symmetry in the
usual sense, the dual symmetries of the theory can simulate the effects of
the former. This leads to a Polyakov-Wiegman type identity which has an
identical structure to the conventional identity.

Before closing this sub-section, it may be useful to highlight some other
aspects of duality which are peculiar to two dimensions, as for instance,
the chiral symmetry. The interpretation of this symmetry with regard to
duality seems, at least to us,
to be a source of some confusion and controversy. As is well known a
scalar field in two dimensions can be decomposed into two chiral pieces,
described by Floreanini Jackiw (FJ) actions \cite{FJ}. 
These actions are sometimes
regarded \cite{PST} as the two dimensional analogues of the duality symmetric 
four dimensional
electromagnetic actions \cite{SS}. Such an interpretation is debatable
since the latter have the $SO(2)$ symmetry (characterised by an internal
index $\alpha$) which is obviously lacking in
the FJ actions. Our analysis, on the other hand, has shown how to
incorporate this symmetry in the two dimensional case. Hence we consider
the actions defined by (\ref{w90}) to be the true analogue of the duality
symmetric electromagnetic actions to be discussed later. 
Moreover, by solving the equations
of motion of the FJ action, it is not possible to recover the second order
free scalar Lagrangean, quite in contrast to the electromagnetic theory 
\cite{SS}.
Nevertheless, since the FJ actions are just the chiral components of the
usual scalar action, these must be soldered to reproduce this result.
But if soldering is possible, such actions must also display the self and
anti-self dual aspects of chiral symmetry. This phenomenon is now explored
along with the soldering process.

The two FJ actions defined in terms of the independent scalar fields
$\phi_+$ and $\phi_-$ are given by,
\be
\label{w220}
{\cal L}^{FJ}_\pm(\phi_\pm)=\pm\dot\phi_\pm\phi_\pm'-\phi_\pm'\phi_\pm'
\ee
whose equations of motion show the self and anti self dual aspects,
\be
\label{w230}
\partial_\mu\phi_\pm=\mp\epsilon_{\mu\nu}\partial^\nu\phi_\pm
\ee
A trivial application of the soldering mechanism leads to,
\br
{\cal L}(\Phi)&=&{\cal L}^{FJ}_+(\phi_+)+{\cal L}^{FJ}_-(\phi_-)+\frac{1}{8}
\Big(J_+(\phi_+)+ J_-(\phi_-)\Big)^2\nonumber\\
&=&\frac{1}{2}\partial_\mu\Phi\partial^\mu\Phi
\label{w240}
\er
where the currents $J_\pm$ and the composite field $\Phi$ are given by,
\br
J_\pm&=&2\Big(\pm\dot\phi_\pm-\phi_\pm'\Big)\nonumber\\
\Phi &=& \phi_+-\phi_-
\label{w250}
\er
Thus the usual scalar action is obtained in terms of the composite field. 
The previous analysis has, however, shown that each of the Lagrangeans
(\ref{w90}) are equivalent to the usual scalar theory. Hence these
Lagrangeans contain both chiralities desribed by the FJ actions
(\ref{w220}). However, in the internal space, ${\cal L}_\pm$ carry the
self and anti self dual solutions, respectively. This clearly illuminates
the ubiquitous role of chirality versus duality in the two dimensional
theories which has been missed in the literature simply because, following
conventional analysis in four dimensions \cite{DT, SS}, 
only one particular duality symmetric Lagrangean ${\cal L}_-$
was imagined to exist. 

\subsection{Coupling to gravity}
It is easy to extend the analysis to include gravity. This is most
economically done by using the language of electrodynamics already
introduced. The Lagrangean for the scalar field coupled to gravity is
given by,
\be
{\cal L}= \frac{1}{2}\sqrt{-g}g^{\mu\nu}F_\mu F_\nu
\label{w260}
\ee
where $F_\mu$ is defined in (\ref{w60b}) and $g=\det g_{\mu\nu}$. Converting
the Lagrangean to its first order form, we obtain,
\be
\label{w270}
{\cal L}=P E
-\frac{1}{2\sqrt{-g}g^{00}}\Big(P^2+B^2\Big)+\frac{g^{01}}{g^{00}} P B
\ee
where the $E$ and $B$ fields are defined in (\ref{w30}) and $P$ is an
auxiliary field. Let us next invoke a change of variables mapping
$(E, B)\rightarrow (E_1, B_1)$ by means of the
$O(2)$ transformation analogous to (\ref{w80}),
and relabel the variable $P$ by $\pm B_2$.
Then the Lagrangean (\ref{w270}) assumes the distinct forms,
\be
\label{w280}
{\cal L}_\pm=
\frac{1}{2}\Bigg[\pm B_\alpha\sigma_{\alpha\beta}^1 E_\beta-\frac{1}
{\sqrt{-g}g^{00}} B_\alpha^2\pm \frac{g^{01}}{g^{00}}
\sigma_{\alpha\beta}^1 B_\alpha B_\beta\Bigg]
\ee
which are duality symmetric under the transformations (\ref{w110}). 
As in the flat metric, there is a swapping between ${\cal L}_+$
and ${\cal L}_-$ if the transformation matrix is $\epsilon_{\alpha\beta}$.
To obtain a duality symmetric action for all 
transformations it is necessary to construct the master action
obtained by soldering the two independent pieces. The dual aspects of the
symmetry that will be soldered are revealed by looking at the equations of
motion following from (\ref{w280}),
\be
\label{w290}
\sqrt{-g}F_\mu^\alpha=\mp g_{\mu\nu}\sigma_1^{\alpha\beta}\mbox{}^*
F^{\nu, \beta} \ee
The result of the soldering process, 
following from our standard techniques, leads to the master Lagrangean,
\be
\label{w310}
{\cal L}=\frac{1}{4}\sqrt{-g}g^{\mu\nu}F_\mu^\alpha F_\nu^\alpha
\ee
where $F_\mu^\alpha$ is defined in terms of the composite field given in
(\ref{w190}).
In the flat space this just reduces to the expression found previously in
(\ref{w180}). It may be pointed out that, originating from this master
action it is possible, by passing to a first order form, to recover the
original pieces. 

To conclude, we show how the FJ action now follows trivially by taking any
one particular form of the two Lagrangeans, say ${\cal L}_+$. To make
contact with the conventional expressions quoted in the literature
\cite{SO}, it is
useful to revert to the scalar field notation, so that,
\be
{\cal L}_+= \frac{1}{2}\Bigg[\phi_1'\dot\phi_2+\phi_2'\dot\phi_1
+2\frac{g^{01}}{g^{00}}\phi_1'\phi_2'-\frac{1}{g^{00}\sqrt{-g}}\phi_\alpha'
\phi_\alpha'\Bigg]
\label{w320}
\ee
This is diagonalised by the following choice of variables,
\br
\phi_1&=&\phi_+ +\phi_-\nonumber\\
\phi_2 &=&\phi_+ - \phi_-
\label{w330}
\er
leading to,
\be
\label{w340}
{\cal L}_+={\cal L}_+^{(+)} (\phi_+, {\cal G}_+)
+{\cal L}_+^{(-)} (\phi_-, {\cal G}_-)
\ee
with,
\br
\label{w350}
{\cal L}_+^{(\pm)} (\phi_\pm, {\cal G}_\pm)&=&\pm\dot\phi_\pm\phi_\pm'
+{\cal G}_\pm\phi_\pm'\phi_\pm'\nonumber\\
{\cal G}_\pm &=&\frac{1}{g^{00}}\Bigg(-\frac{1}{\sqrt{-g}}\pm g^{01}\Bigg)
\er
These are the usual FJ actions in curved space as given in \cite{SO}. Such
a structure was suggested by gauging the conformal symmetry of the free
scalar field and then confirmed by checking the classical invariance under
gauge and affine transformations \cite{SO}. Here we have derived this
result directly from the action of the scalar field minimally coupled to
gravity. 

Observe that the explicit diagonalisation carried out in (\ref{w320}) for
two dimensions is actually a
specific feature of $4k+2$ dimensions. This is related to the basic identity
(\ref{i1}) governing the dual operation. 
If, however, one works with the master (soldered)
Lagrangean, then diagonalisation is possible in either $D=4k+2$ or
$D=4k$ dimensions since the corresponding identity (\ref{i3})
always has the correct signature.

\section{The Electromagnetic Duality}

Exploiting the ideas elaborated in the previous sections, it is straightforward
to implement duality in the electromagnetic theory. Let us start with
the usual Maxwell Lagrangean,
\be
{\cal L}=-\frac{1}{4}F_{\mu\nu}F^{\mu\nu}
\label{m10}
\ee
which is expressed in terms of the electric and magnetic fields
as,\footnote{Bold face letters denote three vectors.}
\be
{\cal L}= \frac{1}{2}\Big(\bf E^2-\bf B^2\Big)
\label{m20}
\ee
where,
\br
E_i&=&-F_{0i}=-\partial_0 A_i+\partial_i A_0\nonumber\\
B_i&=&\epsilon_{ijk}\partial_j A_k
\label{m30}
\er
The following duality transformation,
\be
\bf E\rightarrow \mp\bf B\,\,\,\,;\,\,\,\,\bf B\rightarrow \pm \bf E
\label{m40}
\ee
is known to preserve the invariance of the full set comprising
Maxwell's equations and the Bianchi identities although the Lagrangean
changes its signature. To have a duality symmetric Lagrangean, we now
know how to proceed in a systematic manner. The Maxwell Lagrangean
is therefore recast in a symmetrised first order form,
\be
{\cal L}=\frac{1}{2}\Big(\bf P.\dot{\bf A}-\dot{\bf P}.\bf A\Big)
-\frac{1}{2}{\bf P}^2-\frac{1}{2}\bf B^2
+A_0\bf\nabla.\bf P
\label{m50}
\ee
Exactly as was done for the harmonic oscillator, a change of
variables is invoked. Once again there are two possibilities
which translate
from the old set $(\bf P, \bf A)$ to the new ones $(\bf A_1, \bf A_2)$.
It is, however, important to recall that the Maxwell theory has a constraint
that is implemented by the Lagrange multiplier $A_0$. The redefined variables
are chosen which solve this constraint so that,
\br
\bf P&\rightarrow& \bf B_2\,\,\,\,;\,\,\,\,\bf A\rightarrow \bf A_1\nonumber\\
\bf P&\rightarrow& \bf B_1\,\,\,\,;\,\,\,\,\bf A\rightarrow \bf A_2
\label{m60}
\er
It is now simple to show that, in terms of the redefined variables, the
original Maxwell Lagrangean takes the form,
\be
{\cal L}_\pm={1\over 2}\left(\pm\bf {\dot A}_\alpha
\epsilon_{\alpha\beta}\bf B_\beta
-\bf B_\alpha\bf B_\alpha\right)
\label{m70}
\ee
Adding a total derivative that would leave the equations of motion unchanged,
this Lagrangean is expressed directly in terms of the electric and magnetic
fields,
\be
{\cal L}_\pm={1\over 2}\left(\pm\bf B_\alpha
\epsilon_{\alpha\beta}\bf E_\beta
-\bf B_\alpha\bf B_\alpha\right)
\label{m70a}
\ee
It is duality symmetric under the full $SO(2)$
transformations mentioned in an earlier context. Note that one of the
above structures (namely, ${\cal L}_-$) was given earlier in \cite{SS}.
Once again, in analogy with the harmonic oscillator example, it is
observed that the transformation (\ref{110b}) involving the $R^-$ matrices 
switches the Lagrangeans ${\cal L}_+$ and ${\cal L}_-$ into one another.
The generators of the $SO(2)$ rotations are given by,
\be
\label{john}
Q^{(\pm)}=\mp\frac{1}{2}\int d^3x\,\, {\bf{A}}^\alpha\,.\,{\bf{B}}^\alpha
\ee
so that,
\be
\label{gt}
{\bf{A}}_\alpha\rightarrow {\bf{A}}'_\alpha=e^{-iQ\theta}{\bf {A}}_\alpha
e^{iQ\theta} 
\ee
This can be easily verified by using the basic brackets following
from the symplectic structure of the theory,
\be
\label{girotti}
\Big [A^i_\alpha(x),
\epsilon^{jkl}\partial^k A^l_\beta(y)\Big]=\pm i\delta^{ij}
\epsilon_{\alpha\beta} \delta({\bf{x}}-{\bf{y}})
\ee
It is useful to digress on the significance of the above analysis. Since
the duality symmetric Lagrangeans have been obtained directly from
the Maxwell Lagrangean, it is redundant to show the equivalence of
the former expressions with the latter, which is an essential
perquisite in other approaches. Furthermore, since classical
equations of motion have not been used at any stage, the purported
equivalence holds at the quantum level. The need for any explicit
demonstration of this fact, which has been the motivation of several
recent papers, becomes, in this analysis, superfluous. 
A related observation is that the usual way of showing the classical
equivalence is to use the
equations of motion to eliminate one component from (\ref{m70}),
thereby leading to the Maxwell Lagrangean in the temporal $A_0=0$ gauge.
This is not surprising since the change of variables leading from the
second to the first order form solved the Gauss law thereby
eliminating the multiplier. Finally, note that there are 
two distinct structures for the duality symmetric
Lagrangeans. These must correspond to the opposite aspects of some
symmetry, which is next unravelled. By looking at the equations of
motion obtained from (\ref{m70}),
\be
\bf {\dot A}_\alpha =
 \pm\epsilon_{\alpha\beta}\bf \nabla \times \bf A_\beta
\label{m80}
\ee 
it is possible to
verify that these are just the 
self and anti-self dual solutions,
\be
F_{\mu\nu}^\alpha=\pm\epsilon^{\alpha\beta}\mbox{}^* F_{\mu\nu}^\beta
\,\,;\,\,\mbox{}^* F_{\mu\nu}^\beta=\frac{1}{2}\epsilon_{\mu\nu\rho\lambda}F^{\rho
\lambda}_\beta
\label{m90}
\ee
obtained by setting $A_0^\alpha=0$.  Recall that in the two dimensional
theory the equation of motion naturally assumed a covariant structure.
Here, on the other hand, the introduction of $A_0^\alpha$ is necessary
since this term gives a vanishing contribution to the Lagrangean. This
feature distinguishes a gauge theory from the non gauge theory discussed
in the two dimensional example.
It may be observed that the opposite aspects
of the dual symmetry are contained in the internal space. Following our
quantum mechanical analogy, the next task is to solder the two Lagrangeans
(\ref{m70}). Consider then the gauging of the following symmetry,
\be
\delta \bf H_\alpha={\bf{h}}_\alpha \,\,\,;\,\,\,\bf H=\bf P, \bf Q
\label{m100}
\ee
where $\bf P$ and $\bf Q$ denote the basic fields in the Lagrangeans
${\cal L}_+$ and ${\cal L}_-$, respectively. The Lagrangeans
transform as,
\be
\delta {\cal L}_\pm=\epsilon_{\alpha\beta} 
\Big(\bf \nabla\times{\bf{h}}_\alpha\Big).\bf J_\beta^\pm
\label{m110}
\ee
with the currents defined by,
\be
\bf J_\alpha^\pm(\bf H)=\Big(\mp\bf \dot{H}_\alpha+\epsilon_{\alpha\beta}
\bf \nabla\times\bf H_\beta\Big)
\label{m120}
\ee
Next, the soldering  field $\bf W_\alpha$ is introduced which transforms as,
\be
\delta\bf W_\alpha =-\epsilon_{\alpha\beta}\bf\nabla\times{\bf{h}}_\beta
\label{m130}
\ee
Following standard steps as outlined previously, the final  Lagrangean
which is invariant under the complete set of transformations (\ref{m100})
and (\ref{m130}) is obtained,
\be
{\cal L}={\cal L}_+({\bf P})+{\cal L}_-({\bf Q})-{\bf W}^\alpha\,.\,\Big(\bf
J_\alpha^+ (\bf P)+ J_\alpha^- (\bf Q)\Big)- {\bf W}_\alpha^2
\label{m130a}
\ee
Eliminating the soldering field by using the equations of motion, the effective
soldered Lagrangean following from (\ref{m130a}) is derived,
\be
{\cal L}=\frac{1}{4}\Bigg(\bf {\dot G}_\alpha.{\dot G}_\alpha
-\bf\nabla\times\bf G_\alpha.\bf\nabla\times\bf G_\alpha\Bigg)
\label{m140}
\ee
where the composite field is given by the  combination,
\be
\bf G_\alpha=\bf P_\alpha-\bf Q_\alpha
\label{m150}
\ee
which is invariant under (\ref{m100}).
It is interesting to note that, reinstating the $G_0^\alpha$ variable, 
this is nothing but the Maxwell Lagrangean
with a doublet of fields,
\be
{\cal L}=-\frac{1}{4} G_{\mu\nu}^\alpha G^{\mu\nu}_\alpha\,\,\,;\,\,\,
G_{\mu\nu}^\alpha=\partial_\mu  G_\nu^\alpha-
\partial_\nu  G_\mu^\alpha
\label{m160}
\ee
In terms of the original $ P$ and $Q$ fields it is once again possible,
like the harmonic oscillator example, to write  a Polyakov-Weigman like
identity,
\br
{\cal L}(P-Q)&=&{\cal L}(P)+{\cal L}(Q)-2 W_{i, \alpha}^+( P)
W_{i, \alpha}^-( Q)\nonumber\\
W_{i, \alpha}^\pm(H) &=&\frac{1}{\sqrt 2}\Big(F_{0i}^\alpha(H)
\pm \epsilon_{ijk}\epsilon_{\alpha\beta}
F_{jk}^\beta(H)\Big)\,\,\,;\,\,\, H=P, Q
\label {m170}
\er
With respect to the gauge transformatins (\ref{m100}), the above identity
shows that a contact term is necessary to restore the gauge invariant action
from two gauge variant forms. This, it may be recalled, is just the basic 
content of the Polyakov-Weigman identity. It is interesting to note that
the ``mass" term appearing in the above identity is composed of parity
preserving pieces $W_{i, \alpha}^\pm$, thanks to the presence of the
compensating $\epsilon$-factor from the internal space.

Following the oscillator example, it is now possible to show that by reducing
(\ref{m160}) to a first order from, we exactly obtain the two types of
the duality symmetric Lagrangeans (\ref{m70a}). This shows the equivalence of
the soldering and reduction  processes.

A particularly illuminating way of rewriting the Lagrangean (\ref{m160}) is,
\br
{\cal L} &=&-\frac{1}{8}\Big( G_{\mu\nu}^\alpha + \epsilon^{\alpha\beta}
\mbox{}^* G_{\mu\nu}^\beta\Big)
\Big( G^{\mu\nu}_\alpha -\epsilon_{\alpha\rho}
\mbox{}^* G^{\mu\nu}_\rho\Big)\nonumber\\
&=&-\frac{1}{8}\Big( G_{\mu\nu}^\alpha + 
\tilde G_{\mu\nu}^\alpha\Big)
\Big( G^{\mu\nu}_\alpha -\tilde G^{\mu\nu}_\alpha\Big)
\label{m180}
\er
where, in the second line, the generalised Hodge dual in the space 
containing the internal index has been used
to explicitly show the soldering of the self and anti self dual
solutions. A similar situation prevailed in the two dimensional analysis.
The above Lagrangean manifestly displays the following duality symmetries,
\be
A_{\mu}^\alpha \rightarrow R_{\alpha\beta}^\pm A_{\mu}^{\beta}
\label{m190}
\ee

\noindent where, without any loss of generality, 
we may denote the composite field,
of which $G_{\mu\nu}$ is a function, by $A$. 
The generator of the $SO(2)$ rotations is now given by,
\be
\label{m191}
Q=\int d{\bf{x}}\,\,
\epsilon^{\alpha\beta}{\bf {\Pi}}^\alpha\,\,.\,\,{\bf {A}}^\beta
\ee

Now observe that the master Lagrangean was
obtained from the soldering of two distinct Lagrangeans (\ref{m70}). The
latter were duality symmetric under both $\bf
A_\alpha\rightarrow\pm\epsilon_{\alpha\beta} \bf A_\beta$, while the
transformations involving the $\sigma_1$ matrix interchanged ${\cal L}_+$
with ${\cal L}_-$. The soldered
Lagrangean is therefore duality symmetric under the transformations 
(\ref{m190}). Furthermore, the discrete transformation related to the
$\sigma_1$ matrix has been lifted to its continuous form $R^-$.
The master Lagrangean, therefore, contains a bigger set of duality
symmetries than (\ref{m70}) and, significantly, is also manifestly
Lorentz invariant. Furthermore, recall 
that under the transformations mapping the field
to its dual, the original 
Maxwell equations are invariant but the  Lagrangean changes its signature.
The corresponding transformation in the $SO(2)$ space is
given by, 
\be
\label{0}
G_{\mu\nu}^\alpha \rightarrow R^+_{\alpha\beta}
\mbox{}^*G_{\mu\nu}^\beta
\ee
which, written in component notation, looks like,
\be
\bf E^\alpha\rightarrow\mp\epsilon^{\alpha\beta}\bf B^\beta\,\,\,;\,\,\,
\bf B^\alpha\rightarrow\pm\epsilon^{\alpha\beta}\bf E^\beta
\label{m200}
\ee
The standard duality symmetric Lagrangean fails to manifest this property.
However, as may be easily checked, the equations of motion obtained from 
the master Lagrangean swap with the corresponding Bianchi identity
while the Lagrangean flips sign. In this manner the original property
of the second order Maxwell Lagrangean is retrieved.
Note furthermore that the master Lagrangean possesses the $\sigma_1$ symmetry
(which is just the discretised version of $ R^-$), a feature expected for two
dimensional theories. A similar phenomenon occurred in the previous
section where the master action in two dimensions revealed the $SO(2)$
symmetry usually associated with four dimensional theories.

\subsection{Coupling to gravity}

To discuss how the effects of  gravity are included, we will proceed as in
the two dimensional example. The starting point is the Maxwell Lagrangean
coupled to gravity,
\be
{\cal
L}=-\frac{1}{4}\sqrt{-g}g^{\mu\alpha}g^{\nu\beta}F_{\mu\nu}F_{\alpha\beta}
\label{g10} 
\ee
From our experience in the usual Maxwell theory we know that an eventual 
change of variables eliminates the Gauss law so that the term involving the
multiplier $A_0$ may be ignored from the outset. Expressing (\ref{g10}) in
terms of its components to separate explicitly the first and second order
terms, we find,
\be
\label{g20}
{\cal L}=\frac{1}{2}\dot A_i M^{ij} \dot A_j + M^i \dot A_i +M
\ee
where,
\br
M^{ij}&=&\sqrt{-g}\Big(g^{0i}g^{0j}- g^{ij}g^{00}\Big)\nonumber\\
M^{i}&=&\sqrt{-g} g^{0k}g^{ji} F_{jk}
\nonumber\\
M&=&\frac{1}{4}\sqrt{-g} g^{ij}g^{km} F_{im}F_{kj}
\label{g30}
\er
Now reducing the Lagrangean to its first order form, we obtain,
\be
\label{g40}
{\cal L}= P^i E_i-\frac{1}{2}P^i M_{ij} P^j -\frac{1}{2}M^i M_{ij} M^j
+P^i M_{ij} M^j +M
\ee
where $\dot A_i$ has been replaced by $E_i$ and 
$M_{ij}$ is the inverse of $M^{ij}$,
\be
\label{g50}
M_{ij}= \frac{-1}{\sqrt{-g}g^{00}} g_{ij}
\ee
with,
\be
\label{g60}
g^{\mu\nu}g_{\nu\lambda}=\delta^\mu_\lambda
\ee
Next, introducing the standard change of variables which solves the Gauss
constraint,
\br
\label{g70}
E_i &\rightarrow & E_i^{(1)}\nonumber\\
P^i&\rightarrow &\pm B^{i(2)}
\er
the Lagrangean (\ref{g40}) is expressed in the desired form,
\br
\label{g80}
{\cal L}_\pm=&\pm & E_i^\alpha\epsilon^{\alpha\beta}B_\beta^i
+\frac{1}{\sqrt{-g}g^{00}}g_{ij} B^i_\alpha B_\beta^j\nonumber\\
&\pm & \frac {g^{0k}}{g^{00}} \epsilon_{ijk}
\epsilon^{\alpha\beta}B_\alpha^i B_\beta^j
\er

Once again there are two duality symmetric actions corresponding to ${\cal
L}_\pm$. The enriched nature of the duality and swapping 
symmetries under a bigger set
of transformations, the constructing of a master Lagrangean from
soldering of ${\cal L}_+$ and ${\cal L}_-$, the corresponding
interpretations, all go through exactly as in the flat metric case.
Incidentally, the structure for ${\cal L}_-$ only was previously given in
\cite {SS}. 

\section{Conclusions}
The present work revealed a unifying structure behind the construction of
the various duality symmetric actions. The essential ingredient was the
conversion of the second order action into a first order form followed by
an appropriate redefinition of variables such that these may be denoted
by  an internal index. The duality naturally occurred in this internal
space. Since the duality symmetric actions were directly derived from the
original action the proof of their equivalence becomes superfluous. This
is otherwise essential where such a derivation is lacking and recourse is
taken to either equations of motion  or some hamiltonian analysis.
Obviously the most simple and fundamental manifestation of the duality
property was in the context of the quantum mechanical harmonic oscillator.
Since a field is interpreted as a collection of an infinite set of such
oscillators, it is indeed expected and not at all surprising that all
these concepts and constructions are {\it almost} carried over entirely  
for field theories. It may be remarked that the extension of the harmonic
oscillator analysis to field theories has proved useful in other contexts
and in this particular case has been really clinching. Furthermore, by
invoking a highly suggestive electromagnetic notation for the harmonic
oscillator analysis, its close correspondence with the field theory
examples was highlighted.

A notable feature of the analyis was the revelation of a whole class of
new symmetries and their interrelations. Different aspects of this feature were
elaborated.  To be precise, it was shown that there are actually two
\footnote {Note that usual discussions of duality symmetry consider only
one of these actions, namely ${\cal L}_-$.}
duality symmetric actions $({\cal L}_\pm)$ 
for the same theory. These actions carry the opposite (self and anti self
dual) aspects of some symmetry and their occurrence  was
essentially tied to the fact that there were two distinct classes in which
the renaming of variables was possible, depending on the signature of the
determinant specifying the proper or improper rotations. 
To discuss further the implications of this pair of duality symmetric
actions it is best to compare with the existing results. This also serves
to put the present work in a proper perspective. It should be mentioned
that the analysis for two and four dimensions are generic for $4k+2$ 
and $4k$ dimensions, respectively.

It is usually observed \cite{DGHT} 
that the invariance of the actions in different
$D$-dimensions is preserved by the following groups,
\be
\label{c1}
{\cal G}_d=Z_2 \,\,\,;\,\,\, D=4k+2
\ee
and,
\be
\label{c2}
{\cal G}_c= SO(2)\,\,\,;\,\,\, D=4k
\ee
which are called the ``duality groups". The $Z_2$ group is a discrete
group with two elements, the trivial identity and the $\sigma_1$ matrix.
Observe an important difference
since in one case this group is continous while in the other it is
discrete. In our exercise this was easily verified by the pair of duality
symmetric actions ${\cal L}_\pm$. The new ingredient is that nontrivial
elements of these
groups are also responsible for the swapping ${\cal L}_+\leftrightarrow {\cal
L}_-$, but in the other dimensions. Thus the ``duality swapping matrices"
$\Sigma_s$ are given by,
\br
\label{c3}
{\Sigma}_s &=&\sigma_1\,\,\,;\,\,\, D=4k\nonumber\\
 &=& \epsilon\,\,\,;\,\,\, D=4k+2
\er

It was next shown that ${\cal L}_\pm$ contained the self and anti-self
dual aspects of some symmetry. Consequently, following the ideas developed
in \cite{ABW, BW}, the two Lagrangeans could be soldered to yield a master
Lagrangean ${\cal L}_m={\cal L}_+\oplus {\cal L}_-$. The master action, in
any dimensions, was manifestly Lorentz or general coordinate invariant
and was also duality symmetric under both the groups mentioned
above. Moreover the process of soldering lifted the discrete group $Z_2
$ to its continuous version. The duality group for the master action
in either dimensionality therefore simplified to,
\be
\label{cm}
{\cal G}= O(2)\,\,\,;\,\,\, D=2k+2
\ee
Thus, at the level of the master action, the fundamental distinction 
between the odd and even $N$-forms gets obiliterated. It ought to be
stated that the  lack of usual Chern Simons terms in $D=4k+2$ dimensions
to act as the generators of duality transformations is compensated by the
presence of a similar term in the internal space. Thanks to this it was
possible to  explicitly construct the symmetry generators for the master
action in either two or four dimensions.

We also showed that the master actions in any dimensions, apart from
being duality symmetric under the $O(2)$ group, were factored, modulo a
normalisation,  as a
product of the self and anti self dual solutions,
\be
\label{1}
{\cal L}=\Big (F^\alpha +\tilde F^\alpha\Big) 
\Big (F^\alpha -\tilde F^\alpha\Big)\,\,\,;\,\,\,D=2k+2
\ee
where the internal index has been explicitly written and the
generalised Hodge operation was defined in (\ref{i2}). The key ingredient
in this construction was 
to provide a general definition of self duality that was applicable
for either odd or even $N$ forms. Self duality was now defined to
include the internal space and was implemented either by the
$\sigma_1$ or the $\epsilon$, depending on the dimensionality. 
This naturally led to the universal structure (\ref{1}).

Some other aspects of the analysis deserve attention. Specifically, the  
novel  duality symmetric actions obtained in two dimensions
revealed the interpolating role between duality and chirality.
Furthermore, certain points concerning the
interpretation of chirality symmetric action as the
analogue of the duality symmetric electromagnetic action in four
dimensions were clarified.
We also recall that the
soldering of actions to obtain a master action was an intrinsically
quantum phenomenon that
could be expressed in terms of an identity relating two ``gauge variant"
actions to a ``gauge invariant" form. The gauge invariance is with regard
to the set of transformations induced for effecting the soldering and has
nothing to  do with the conventional gauge transformations. In fact the
important thing is that the distinct actions must possess the self and
anti self dual aspects of some symmetry which are being soldered. The
identities obtained in this way are effectively a
generalisation of the usual Polyakov Weigman identity. We conclude by stressing
the practical nature of our approach to duality which can be extended to
other theories.

\bigskip
\bigskip

{\bf Acknowledgements}
\bigskip

One of the authors (R.B) would like to thank the CNPq for providing
financial support and the members of the Physics Deptt., UFRJ, for their
kind hospitality. C.W. wants to thank Prof. J. Mignaco for many
interesting discussions. He also acknowledges partial financial support
from CNPq, FUJB and FINEP.

\end{document}